\documentclass[a4paper]{article}
\usepackage{ifthen}
\usepackage{epsfig}
\usepackage{amssymb}
\usepackage{graphicx}
\usepackage{amsmath}
\usepackage[center]{subfigure}
\renewcommand{\text}[1]{%
\ifthenelse{\equal{#1}{fB}}{f_B}{}% 
\ifthenelse{\equal{#1}{fD}}{f_D}{}%
\ifthenelse{\equal{#1}{mc}}{m_c}{}%
\ifthenelse{\equal{#1}{mB}}{m_B}{}%
\ifthenelse{\equal{#1}{mD}}{m_D}{}%
\ifthenelse{\equal{#1}{M2}}{M^2}{}%
\ifthenelse{\equal{#1}{fDst}}{f_{D^*}}{}%
\ifthenelse{\equal{#1}{mDst}}{m_{D^*}}{}%
\ifthenelse{\equal{#1}{barxi}}{\bar\sigma}{}%
\ifthenelse{\equal{#1}{uu}}{u}{}%
\ifthenelse{\equal{#1}{xi0}}{\sigma_0}{}%
\ifthenelse{\equal{#1}{s0D}}{s_0^D}{}%
\ifthenelse{\equal{#1}{ubar}}{\bar{\sigma}}{}%
\ifthenelse{\equal{#1}{vv}}{v}{}
\ifthenelse{\equal{#1}{phiBp}}{\phi^B_+}{}%
\ifthenelse{\equal{#1}{PhiBp}}{\phi^B_+}{}%
\ifthenelse{\equal{#1}{phiBmin}}{\phi^B_-}{}%
\ifthenelse{\equal{#1}{PhiBmin}}{\phi^B_-}{}%
\ifthenelse{\equal{#1}{PhiBpm}}{\overline\Phi^B_{\pm}}{}%
\ifthenelse{\equal{#1}{PhiBpmin}}{\overline\Phi^B_{\pm}}{}%
\ifthenelse{\equal{#1}{ppsiV}}{\Psi_V^B}{}%
\ifthenelse{\equal{#1}{ppsiA}}{(\Psi_A^B}{}%
\ifthenelse{\equal{#1}{bbarXA}}{\Psi_V^B}{}%
\ifthenelse{\equal{#1}{bbarYA}}{\overline Y_A^B}{}%
\ifthenelse{\equal{#1}{lb}}{\lambda_B}{}%
}
\newcommand{\Dst}{D^{(*)}}
\newcommand{\bom}{\overline{\omega}}

\newcommand{\ba}{\begin{eqnarray}}
\newcommand{\ea}{\end{eqnarray}}
\newcommand{\be}{\begin{equation}}
\newcommand{\ee}{\end{equation}}
\newcommand{\DS}[1]{/\!\!\!#1}

\makeatletter
\def\fmslash{\@ifnextchar[{\fmsl@sh}{\fmsl@sh[0mu]}}
\def\fmsl@sh[#1]#2{%
  \mathchoice
    {\@fmsl@sh\displaystyle{#1}{#2}}%
    {\@fmsl@sh\textstyle{#1}{#2}}%
    {\@fmsl@sh\scriptstyle{#1}{#2}}%
    {\@fmsl@sh\scriptscriptstyle{#1}{#2}}}
\def\@fmsl@sh#1#2#3{\m@th\ooalign{$\hfil#1\mkern#2/\hfil$\crcr$#1#3$}}
\makeatother
\newcommand{\ps}[0]{\fmslash{p\,}\:\!}

\begin{document}
\begin{titlepage}
%\begin{flushright}
%SI-HEP-2008-13\\
%\end{flushright}
\vfill
\begin{center}
{\Large\bf 
$B\!\to\!\! D^{(*)}$ Form Factors \\[2mm] from QCD Light-Cone Sum Rules}\\[2cm]
{\large\bf  
S.~Faller\,$^{(a,b)}$, A.~Khodjamirian\,$^{(a)}$, 
Ch.~Klein\,$^{(a)}$ and Th.~Mannel\,$^{(a)}$}\\[0.5cm]
{\it  $^{(a)}$\,Theoretische Physik 1, Fachbereich Physik,
Universit\"at Siegen,\\ D-57068 Siegen, Germany }\\[2mm]
{\it $^{(b)}$ \,Theory Division, Department of Physics,
CERN,\\ CH-1211 Geneva 23, Switzerland }
\end{center}
\vfill
\begin{abstract}
We derive new QCD sum rules 
for $B\to D$ and $B\to D^*$ form factors. 
The underlying correlation functions  
are expanded near the light-cone in terms of $B$-meson 
distribution amplitudes defined in HQET,
whereas the $c$-quark mass is kept finite. The leading-order
contributions of two- and three-particle distribution amplitudes
are taken into account. From the resulting light-cone sum rules 
we calculate all $B\to \Dst $ form factors 
in the region of small momentum transfer (maximal recoil). 
In the infinite heavy-quark mass limit the sum rules reduce 
to a single expression for the Isgur-Wise function.
We compare our predictions with the 
form factors extracted from experimental $B\to \Dst l \nu_l$ decay
rates fitted to dispersive parameterizations. 

\end{abstract}
\vfill
\end{titlepage}

\section{Introduction}
The hadronic form factors of $B\!\to\! D,D^*$ transitions are used 
to extract the CKM parameter $V_{cb}$ from the 
measurements of the semileptonic $B\to \Dst l \nu_l$ decay rates.
These form factors were among the first and most important 
applications of the heavy quark symmetry \cite{SV,IW}
and heavy quark effective theory (HQET)\cite{hqet}
(for reviews see \cite{Neubert,manohar,Uralt}).

In the heavy-quark limit, all $B\!\to\! \Dst $ form factors are 
expressed via the 
Isgur-Wise (IW) function  $\xi(w)$ of the velocity transfer 
$w=v\cdot v'$ in  the $B(v)\!\to\! \Dst(v')$ transition. 
The gluon radiative and 
inverse heavy-mass corrections are  well understood 
within heavy-quark expansion and HQET
(see e.g., \cite{Neubert,Uralt,CzM}), in particular 
at the zero recoil ($w=1$) point. The
$B\!\to\!\Dst$ form factors are also being calculated
in lattice QCD \cite{latt1,bernard,latt2}. 
Beyond the zero-recoil  point, at $w>1$, one usually 
parameterizes these form factors
\cite{BGL,CLN}, employing conformal
mapping and dispersive bounds based on analyticity and unitarity. 
The data on $B\to \Dst l \nu_l$,
including the most recent measurements \cite{BaBar1,BaBar2}
are fitted to these parameterizations. 

For a better theoretical description of $B\to \Dst l 
\nu_l $ transitions in the whole kinematical region and for a 
quantitative assessment 
of $1/m_Q$ corrections it is desirable to perform alternative 
calculations of the form factors within full QCD, with finite 
heavy quark masses, at least, with a finite 
$c$-quark mass. 
Previously, $B\to \Dst$ form factors 
were calculated from QCD sum rules for 
three-point correlation
functions with finite $b$- and $c$-quark masses
\cite{QCDSRBD,BallBD}. These calculations employ  
the local operator-product 
expansion (OPE) and include nonperturbative 
effects in the form of quark and gluon condensates.
Based on double dispersion relations,
the three-point sum rules  are
quite sensitive to  the choice of the quark-hadron duality  region. 
The heavy-quark limit of three-point sum rules 
reproduces a universal IW function and
reveals noticeable corrections from finite quark masses
(see e.g.,\cite{BallBD}) .
A direct calculation of the IW function 
from the sum rules in HQET is also possible
\cite{Neubertetal,Rad,Neub92,BS}, including $O(\alpha_s)$ corrections 
\cite{BBG}.

A well known  alternative sum rule approach to hadronic form factors  
relies on the OPE near the light-cone \cite{lcsr} and employs the 
light-cone distribution amplitudes of hadrons.
This approach has been successfully applied to heavy-light 
form factors (some recent results can be found in 
\cite{BZBpi,KMO1,KMO2,DFH,DKMMO}). It is timely to develop 
a similar technique also for the $B\to \Dst $ form factors. 

In this paper we apply the recently suggested version 
of QCD light-cone sum rules \cite{KMO1,KMO2},
in which the set of $B$-meson distribution amplitudes 
(DA's) serves as a universal nonperturbative input 
(in \cite{DFH} a similar approach was used in the framework of SCET). 
We keep the $c$-quark mass finite
and employ the quark-hadron duality approximation in 
the $\Dst$ channel of the correlation function. 
The on-shell $B$ meson is treated in HQET, 
to allow the expansion in DA's. As discussed below, the
light-cone expansion is applicable  
in the region of maximal recoil. We obtain 
predictions  for all $B\to \Dst $ form factors 
in this region  and compare the results 
with the experimental data on semileptonic decay rates.
We also derive the infinite heavy-quark mass limit of the new sum rules.

The plan of the paper is as follows.
In section 2 we introduce the correlation function and 
derive the sum rules for the form factors. 
In section 3 we switch to the form factors adapted to 
heavy-quark symmetry and discuss the heavy-mass limit of the sum rules.
Section 4 is devoted to numerical results. In section 5 we conclude.
The appendix contains definitions of $B$-meson DA's and the 
bulky expressions for three-particle  contributions to sum rules. 

\section{Correlation function and sum rules}

Following \cite{KMO2}, we consider the correlation function 
of two quark currents taken between the vacuum and 
the on-shell $\bar{B}$-meson state:  
\begin{equation}
F_{a\mu}^{(B)}(p,q)= i\int d^4x ~e^{i p\cdot x}
\langle 0|T\left\{\bar{d}(x)\Gamma_a c(x), 
\bar{c}(0)\gamma_\mu(1-\gamma_5) b(0)\right\}|\bar{B}(p_B)\rangle\,,
\label{eq-corr}
\end{equation}
where the weak $b\to c$ current  is correlated with the 
$\bar{d}(x)\Gamma_a c(x)$ current. The latter interpolates 
the pseudoscalar $D$-meson ($\Gamma_a=m_ci\gamma_5$) or vector $D^*$-meson 
($\Gamma_a=\gamma_\nu$). For definiteness, we choose the
$\bar{B}_d\to D^{(*)+}$ transition, equivalent to 
$\bar{B}_u\to D^{(*)0}$  in the isospin symmetry
limit. The external momenta of the weak and interpolating 
currents are $q$ and $p$, respectively,
with the $B$ meson momentum being on-shell, $p_B^2=(p+q)^2=m_B^2$ . 

The correlation function 
(\ref{eq-corr}) is related to the form factors of our interest 
via  the hadronic dispersion relation in the channel 
of the charmed meson:
\be
F_{a\mu}^{(B)}(p,q) =\frac{\langle 0|\bar{d}\Gamma_a c|\Dst(p)\rangle\langle
  \Dst(p)|\bar{c}\gamma_\mu(1-\gamma_5) b|\bar{B}(p+q)\rangle
}{m_{\Dst}^2-p^2}+...\,,
\label{eq-disp}
\ee
where the $\Dst$-meson pole term
is shown explicitly, and ellipses indicate 
the contributions of excited and continuum states. The r.h.s.
of Eq.~(\ref{eq-disp}) contains the decay constant:
\be
\langle 0|\bar{d}m_ci\gamma_5 c|D(p)\rangle
=m_D^2 f_D, 
\label{decconst1}
\ee
or
\be
\langle 0|\bar{d}\gamma_\nu c|D^*(p,\epsilon)\rangle=
\epsilon_\nu m_{D^*}f_{D^*}\,,
\label{decconst2}
\ee
(where $\epsilon$ is the polarization vector of $D^*$)
and the $B\to \Dst $ transition  matrix element.
The latter is determined by the form factors 
for which we use the standard definitions:
\begin{eqnarray}
&&\langle D(p)|\bar{c}\gamma_\mu b| \bar{B}(p+q)\rangle
=2p_\mu f^+_{BD}(q^2)+q_\mu \left[f^+_{BD}(q^2)+f^-_{BD}(q^2)\right]\,,
\label{eq-formfBD}
\end{eqnarray} 
and  
\begin{eqnarray}
\langle D^{*}(p,\epsilon)|\bar{c}\gamma_\mu(1-\gamma_5) b| \bar{B}(p+q)\rangle
=-i\epsilon_\mu^*(m_B+m_{D^*})A_1^{BD^*}(q^2)
\nonumber\\
+i(2p+q)_\mu(\epsilon^*q)\dfrac{A_2^{BD^*}(q^2)}{m_B+m_D{^*}}+
iq_\mu(\epsilon^* q)\dfrac{2m_{D^*}}{q^2}\Big(A_3^{BD^*}(q^2)-A_0^{BD^*}(q^2)\Big)
\nonumber\\
+\epsilon_{\mu\nu\rho\sigma} 
\epsilon^{*\nu}q^\rho p^\sigma\dfrac{2 V^{BD^*}(q^2)}{m_B+m_{D^*}}\,,
\label{eq-formfBDst}
\end{eqnarray}
where $A_0^{BD^*}(0)=A_3^{BD^*}(0)$ and 
$$2m_D^* A_3^{BD^*}(q^2)= (m_B+m_{D^*})A_1^{BD^*}(q^2)-
(m_B-m_{D^*})A_2^{BD^*}(q^2)\,.$$

The sum rule derivation follows the procedure  
similar to the  one applied in \cite{KMO2}. Instead of the 
virtual light quark now the $c$ quark propagates in the correlation
function. The calculation is performed 
in terms of $B$-meson DA's defined in HQET,
hence the correlation function (\ref{eq-corr})
has to be expanded in the limit of large $m_b$:
\begin{equation}
F_{a\mu}^{(B)}(p,q)=\tilde{F}_{a\mu}^{(B_v)}(p,q')
+O(1/m_b)\,,
\label{eq-HQET}
\end{equation}
where the limiting correlation function is
\begin{equation}
\tilde{F}_{a\mu}^{(B_v)}(p,q')
= i\int d^4x ~e^{i p\cdot x}
\langle 0|T\left\{\bar{d}(x)\Gamma_a c(x), 
\bar{c}(0)\gamma_\mu(1-\gamma_5)h_v(0)\right\}|\bar{B}_v\rangle\,. 
\label{eq-HQETcorr}
\end{equation}
and each term of this expansion retains dependence on finite $m_c$.
In  the above, 
the four-momentum of the $B$-meson state is redefined as $p_B=m_bv +k$, 
where $v$ is the four-velocity of $B$, $k$ is the residual 
momentum, and the relativistic normalization of the state 
$|\bar{B}(p_B)\rangle=|\bar{B}_v\rangle $ (up to $1/m_b$ corrections) is retained. 
In addition, the $b$-quark field 
is substituted by the effective field, using 
$b(x) = h_v(x)e^{-im_bvx}+O(1/m_b)$,
and the four-momentum transfer $q$ 
is redefined by separating the ``static'' part of it: 
$q=m_b v+q'$. In what follows, the initial correlation function
(\ref{eq-corr})
is calculated  in the approximation (\ref{eq-HQETcorr}). 
\footnote{The 
subleading $O(1/m_b)$ correlation functions  can in principle be 
obtained  if one expands both quark-current operator and 
$B$ state in powers of $1/m_b$.} 
Note that (\ref{eq-HQETcorr}) does not depend on $m_b$ since the 
external momentum scales $p$ and $q'$ are generic and do not scale with 
the heavy quark mass. 

Before turning to the calculation, it is important 
to convince  oneself that the light-cone dominance
is valid for off-shell external momenta $p$ and $q$, that is, 
if $p^2$ and $q^2$ are far below the 
hadronic thresholds in the channels of
$\bar{d}\Gamma_a c$ and $\bar{c}\gamma_\mu(1-\gamma_5)b$ 
currents, respectively.  
To demonstrate that, we 
can use the same line of arguments as in \cite{KMO2}.
For simplicity, we consider the rest frame $v=(1,0,0,0)$, where, in first 
approximation, $m_B= m_b+\bar \Lambda$, so that  
$k_0\sim \bar{\Lambda}$. In addition, it is also convenient to rescale 
the $c$-quark field by introducing an effective field  
$h'_v(x)=c(x)e^{-im_cvx}$ (with the same velocity). 
Simultaneously, the external four-momenta are 
redefined: $p=m_cv+\tilde{p}$, $q'=-m_cv+\tilde{q}$, 
separating the parts proportional to the velocity $v$, 
so that 
$q=(m_b-m_c)v+\tilde{q}$, and $\tilde{p}+\tilde{q}=k$.
Note that the last redefinitions do not necessarily mean
that we will use HQET also for the virtual 
$c$-quark field.
It is done only  in order to 
decouple the $c$-quark mass scale. Indeed, we now 
arrive at a modified correlation function 
\begin{equation}
\tilde{F}_{a\mu}^{(B_v)}(\tilde{p},\tilde{q})
= i\int d^4x ~e^{i \tilde{p}\cdot x}
\langle 0|T\left\{\bar{d}(x)\Gamma_a h'_v(x), 
\bar{h}_v'(0)\gamma_\mu(1-\gamma_5)h_v(0)\right\}|\bar{B}_v\rangle 
\label{eq-HQETcorr1}
\end{equation}
of  two effective currents 
with the external momenta $\tilde{p}$, $\tilde{q}$. 
This correlation function does not explicitly depend
on both $b$- and $c$- quark masses and contains  
only the scales associated with either effective or light-quark 
degrees of freedom.

We assume that both rescaled four-momenta 
are  spacelike and their squares are sufficiently large: 
\be 
P^2, |\tilde{q}^2|\gg \Lambda_{QCD}^2,
\bar{\Lambda}^2\,,
\label{eq-scale2}
\ee
where $P^2= -\tilde p^2$. Furthermore, the difference 
between the virtualities is also kept large, so that the ratio 
\be  
\zeta=\frac{2 \tilde p\cdot k }{P^2}\sim \frac{|\tilde{q}^2|-P^2}{P^2}
\sim 1\,.  
\label{eq-xi}
\ee
With these two conditions fulfilled,
the region of small $x^2\leq 1/P^2$  
dominates in the integral in (\ref{eq-HQETcorr1}),
in full analogy with 
the $\gamma^*(\tilde p)\gamma^*(\tilde{q})\to \pi^0
(\tilde{p}+\tilde{q})$ 
transition amplitude,
for which a detailed proof of the light-cone dominance 
can be found e.g., in \cite{CK}. 
Thus, the choice of large $P^2$ and $\zeta\sim 1$ 
enables the validity of light-cone OPE. 
In terms of the initial external momenta
$p$ and $q$, one now has 
\ba
p^2&=&m_c^2-\zeta m_cP^2/\bar{\Lambda}-P^2\,,
\nonumber
\\
q^2&=&(m_b-m_c)^2 -(m_b-m_c)P^2\zeta(1+\zeta)/\bar{\Lambda}-
P^2(1+\zeta)\,,
\label{eq-tau}
\ea
taking into account 
that $\tilde{p}_0=\zeta P^2/(2\bar{\Lambda})$ in the rest frame. 
Note that  
the external momentum squared $p^2$  
in the charmed meson channel has to be shifted below the threshold 
$m_{\Dst}^2\sim m_c^2$ by an interval $\sim m_c\chi$. 
The scale 
$\chi\sim  P^2/\bar{\Lambda}\gg\bar{\Lambda},\Lambda_{QCD}$
is large in terms of$\Lambda_{QCD}$ , but in general 
independent of the heavy quark masses.  
The situation here is quite similar to the 
correlation function used to derive LCSR  
for $B\to \pi$ form factors with pion DA's 
(see e.g.,\cite{BZBpi,DKMMO}), 
in which case the light-cone dominance 
is guaranteed by off-shell external momenta.     
Importantly, the second condition in (\ref{eq-tau}) 
tells us that OPE is only  applicable
sufficiently far from the zero recoil (maximal $q^2$) point 
$q^2=(m_B-m_\Dst)^2\sim (m_b-m_c)^2$ of the $B\to \Dst$ 
transition. In practice, LCSR will be applied  
at $q^2\sim 0$, near the maximal recoil. Solving the 
second equation in   
(\ref{eq-tau}) for $q^2=0$ we obtain $P^2\sim \bar{\Lambda} 
(m_b-m_c)\ll m_b^2$, however, the components 
of the external momenta reach
the order of magnitude of the heavy-quark mass  scale.

Returning to the correlation function (\ref{eq-HQETcorr}), we
calculate the leading order (LO) contributions
of two- and three-particle $B$-meson DA's. The corresponding
diagrams are depicted in Figs. 1a and 1b, respectively. 
We use the
$c$-quark propagator near the light-cone,
including the one-gluon part \cite{BB}:
\ba
S_{c}(x,0)=-i\langle 0 |T\{c(x)\bar{c}(0)\}|0 \rangle=
\!\int\!\dfrac{d^4p}{(2\pi)^4}e^{-ipx}
\Bigg\{\dfrac{\ps+m_{c}}{p^2-m_{c}^2}
\nonumber\\
+\!\int_0^1\!d\alpha\,G_{\mu\nu}(\alpha x)
\left[\dfrac{\alpha x^\mu
    \gamma^\nu\,}{p^2-m_{c}^2}
-\dfrac{(\ps+m_{c})\sigma^{\mu\nu}}{2(p^2-m_{c}^2)^2}
\right]\Bigg\}\,,
\label{eq-prop}
\ea
where $G_{\mu\nu}=g_sG^a_{\mu\nu}(\lambda^a/2)$.  
Calculating the correlation function, we confine 
ourselves by the zeroth order in $\alpha_s$, 
hence the $O(\alpha_s)$ differences between 
various $c$-quark mass definitions are beyond our
accuracy. Generally, since there is a highly  
virtual $c$-quark in the correlation function, 
(and in anticipation of future $O(\alpha_s)$ corrections to 
LCSR), the most natural choice is the $\overline{MS}$ 
mass, which we adopt in numerical calculations.

After contracting $c$-quark fields 
and substituting the propagator (\ref{eq-prop}) 
the correlation function is expressed in terms 
of two- and three-particle DA's of the $B$-meson.
Their definitions are given in  the Appendix,
where we also specify  
the adopted  exponential model of two-particle DA's 
suggested in \cite{GN} and the corresponding set of three-particle
DA's derived in \cite{KMO2} from QCD sum rules in HQET.
%%%%%%%%%fig1: Diagrams 
\begin{figure}[t]
\includegraphics[width=14cm]{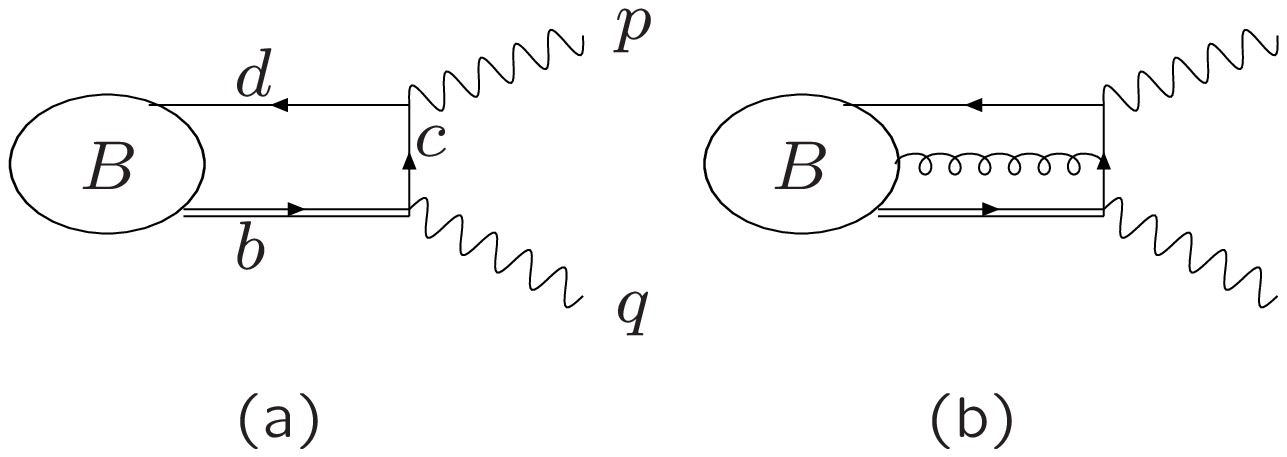}\\
\vspace{-1cm}
\caption{ \it Diagrams corresponding to 
the contributions of 
(a) two-particle and (b) three-particle $B$-meson DA's 
to the correlation function
(\ref{eq-corr}); 
Curly (wavy) lines denote gluons (external currents).}
\label{fig-diags}
\end{figure}
%%%%%%%%%%%%%%%%%%%%%%%%%%%%%%%%%%%

The result for the correlation function  is 
equated to the hadronic representation (\ref{eq-disp}).
Each independent Lorentz-structure in this 
equation provides a sum rule relation for a certain 
form factor or  a combination of form factors. 
In the correlation function for $B\to D$ form factors we take 
the coefficients at $p_\mu$ and $q_\mu$ to obtain  the  sum rules
for the form factors $f^ +_{BD}$ and   $f^+_{BD}+f^-_{BD}$,
respectively. In the $B\to D^*$ case, we choose 
the kinematical structures 
$\epsilon_{\mu\nu\rho\sigma}q^\rho p^\sigma$,
$g_{\mu\nu}$ and $p_\mu q_\nu $ for the form factors 
$V^{BD^*}$, $A_1 ^{BD^*}$ and $A_2 ^{BD^*}$, respectively. 
To obtain the sum rule for the remaining 
combination of form factors $ A_3 ^{BD^*}-A_0 ^{BD^*}$, 
the sum rule for the invariant amplitude multiplying 
 $q_\mu q_\nu$ in the correlation function
has to be derived, from which the sum rule 
for $A_2 ^{BD^*}$ has to be subtracted. 
The further derivation of the sum rules does not differ 
from the procedure explained in \cite{KMO2} and we will not
repeat the details here.

First, we present the sum rules for the two 
$B\to D$  form factors: 

%%%%%%%%%%%%%%%%%%%%%%%%%%%%% f+BD %%%%%%%%%%%%%%%%%%
\begin{multline}
f^+_{BD}(q^2)=
\frac{\text{fB}m_Bm_c}{2\text{fD}m_D^2}
\Bigg\{ 
\int\limits_0^{\omega_0(q^2,\text{s0D})} d\omega 
\exp\left(\frac{-s(\omega,q^2)+m_D^2}{\text{M2}}\right) \\
\times \left[
\frac{m_c(\bom+m_c)}{\bom^2+m_c^2-q^2}
\text{phiBmin}(\omega) +
%%%%%%%%%%%%%%%%%%%%%%%%%%%%%%%%%%%%%%
(\bom+m_c)\left(\frac{1}{\bom}
-\frac{m_c}{
\bom^2+m_c^2-q^2}\right) 
\text{phiBp}(\omega)
%%%%%%%%%%%%%%%%%%%%%%%%%%%%%%%%%%%%%%%%%%%%%
%%
\right.
\\
\left.
-\left(\frac{1}{\bom}+ \frac{m_c(\bom^2+2m_c\bom-m_c^2+q^2)}{
\left(\bom^2+m_c^2-q^2\right)^2}\right)\text{PhiBpm}(\omega) \right] 
+\Delta f^+_{BD}(q^2,s_0^D,M^2)
\Bigg\} \,, 
%%%%%%%%%%%%%%%%%%%%%%%%%%%%%%%%%%%%%%%%%%%%%%%%%%%%%%%%%%%
\label{eq:fplBD}
\end{multline}

%%%%%%%%%%%%%%%%%%%%%%%%%%%%%%% f+BP +f-BP %%%%%%%%%%%%%%%%
\begin{multline}
f^+_{BD}(q^2)+f^-_{BD}(q^2)=
-\frac{\text{fB}m_Bm_c}{ \text{fD}m_D^2} 
\Bigg\{\int\limits_0^{\omega_0(q^2,\text{s0D})} \, d\omega 
\exp\left(\frac{-s(\omega,q^2)+m_D^2}{\text{M2}}\right) \\
%%%%
 \times\left[\frac{m_c(\omega-m_c)}{\bom^2+m_c^2-q^2}
\text{phiBmin}(\omega)
+(\omega-m_c)\left(
\frac{1}{\bom}
-\frac{m_c}{\bom^2+m_c^2-q^2}\right)  
\text{phiBp}(\omega)
\right.\\
\left.
+\left(\frac{1}{\bom}-
\frac{m_c\left(m_B^2-\omega^2-2m_c\bom+m_c^2-q^2\right)}{\left(\bom^2 
+m^2_c-q^2\right)^2}
\right)\text{PhiBpm}(\omega)\right]\\
    +\Delta f^{\pm}_{BD}(q^2,s_0^D,M^2)
\Bigg \} \,, 
%%%%%%%%%%%%%%%%%%%%%%%%%%%%%%%%%%%%%%%%%
\label{eq:fpmBD}
\end{multline}
where the following notations are used: $\bom=m_B-\omega$, 
\be
\text{PhiBpm}(\omega )=\int\limits_0^{\omega} d\tau \Big(\phi^B_+(\tau )-\phi^B_-(\tau)\Big)
\nonumber
\ee 
and 
\be
s(\omega,q^2)=m_B\omega+
\frac{m_c^2m_B-q^2\omega }{\bom}\,.
\nonumber
\ee
The threshold $s_0^{D^{(*)}}$  in the charmed meson channel
transforms into the upper limit of the $\omega$-integration:  
\be
\omega_0(q^2,s_0^{D^{(*)}})=
\frac{\text{mB}^2-q^2+s_0^{D^{(*)}}-\sqrt{4 \left(m_c^2-s_0^{D^{(*)}}\right)
    \text{mB}^2+\left(\text{mB}^2-q^2+s_0^{D^{(*)}}\right)^2}}{2 \text{mB}}\,.
\nonumber
\ee
In the above sum rules, 
$\Delta f_{BD}^+$ and $\Delta f_{BD}^{\pm}$  denote the contributions
of three-particle DA's calculated from the diagram in Fig.1b.
Their bulky expressions are presented in the Appendix.
Note that the heavy-mass scale $m_B$ and related 
$m_c/m_B$ terms in the sum rules originate from  
the propagator of the virtual $c$ quark. The latter  depends on the 
external momenta $q$ and $p$ which,  as explained above,   
satisfy (\ref{eq-tau}).

The analogous sum rules for the three most 
important $B\to D^*$ form factors $V,A_1,A_2$ are 
simply reproduced from the sum rules for the heavy-light 
$B\to K^*$ form factors obtained in \cite{KMO2},
making a replacement $m_s\to m_c$ and switching to 
the same notations as in (\ref{eq:fplBD}),
(\ref{eq:fpmBD}):  

%%%%%%%%%%%%%%%%%%%%%%%%%%%%%%%%%%%%%%%%%%%%%%%%%%%%%

\begin{multline}
V^{BD^*}(q^2)=\frac{\text{fB} \text{mB}}{2 \text{fDst} \text{mDst}} 
(\text{mB}+\text{mDst})
 \Bigg\{\int\limits_0^{\omega_0(q^2,s_0^{D^*})}\, d\omega  
\exp\left(\frac{-s(\omega,q^2)+m_{D^*}^2}{\text{M2}}\right) \\
%%%%
\times \left[\frac{m_c}{\bom^2+m_c^2-q^2}\text{phiBmin}(\omega)
+\left(\frac{1}{\bom}-\frac{ m_c
    }{\bom^2+m_c^2-q^2}\right) \text{phiBp}(\omega )
\right.\\
\left.
-\frac{2 
m_c\bom }{\left(\bom^2+m_c^2-q^2\right)^2}
\text{PhiBpm}(\omega)\right]
+\Delta V^{BD^*}(q^2,s_0^{D^*},M^2)\Bigg \} \,,  
\label{eq:VBDst}
\end{multline}

\begin{multline}
A_1^{BD^*}(q^2)=
\frac{\text{fB} \text{mB}^2}{2 \text{fDst} \text{mDst}
    (\text{mB}+\text{mDst})} 
\Bigg\{\int\limits_0^{\omega_0(q^2,s_0^{D^*})}\, d\omega
\exp\left(\frac{-s(\omega,q^2)+m_{D^*}^2}{\text{M2}}\right) \\
\times
\left[\dfrac{(\bom+m_c)^2-q^2}{\bom^2}\left
\{\dfrac{m_c\bom}{\bom^2+m_c^2-q^2}\text{PhiBmin}(\omega )
%%
%\right.\right.
% \\
%\left.\left.
%%
+\left(1-\dfrac{m_c\bom}{\bom^2+m_c^2-q^2}\right)\text{PhiBp}(\omega)\right\}
\right.
 \\
\left.
-\dfrac{4\bom m_c^2}{\left(\bom^2+m_c^2-q^2\right)^2}
\text{PhiBpm}(\omega)\right]
+\Delta A_1^{BD^*}(q^2,s_0^{D^*},M^2)
\Bigg \} \,,   
\label{eq:A1BDst}
\end{multline}

%%%%%%%%%%%%%%%%%%%%%%%%%%%%%%%%%%%%%%%%%%%%%%%%%%
\begin{multline}
A_2^{BD^*}(q^2)=\frac{\text{fB}}{2 \text{fDst} m_{D^*}}(m_B+m_{D^*}) 
\Bigg\{\int\limits_0^{\omega_0 (q^2,s_0^{D^*})} \, d\omega
\exp\left(\frac{-s(\omega,q^2)+m_{D^*}^2}{\text{M2}}\right) 
\\
\times\left[\frac{(m_c\text{mB}-2 \bom \omega)}{\bom^2+m_c^2-q^2}
\text{phiBmin}(\omega)
+\left(1\,-\,\frac{\omega}{\bom}-
\frac{(m_cm_B-2 \omega \bom)}{\bom^2+m_c^2-q^2}\right)\text{phiBp}(\omega)
\right.
\\
\left.
%%%
-2\left(
\frac{\bom (m_cm_B-2 \omega\bom )}{\left( \bom^2 +m_c^2-q^2\right)^2}+
\frac{ (\omega -\bom) }{\bom^2+m_c^2-q^2}\right)
\text{PhiBpm}(\omega)\right]+
\Delta A_2^{BD^*}(q^2,s_0^{D^*},M^2)
\Bigg \} \,,   
\label{eq:A2BV}
\end{multline}
%%%%%%%%%%%%%%%%%%%%%%%%%%%%%%%%%%%%%%%%%%%%%%%%%%%%%
Finally, we present a new sum rule for the 
remaining combination of $B\to D^*$ form factors:

%%%%%%%%%%%%%%%%%%%%
\begin{multline}
A_3^{BD^*}\left(q^2\right)-A_0^{BD^*}\left(q^2\right)=
\frac{f_B\,q^2}{4f_{D^*}m_{D^*}^2}\Bigg 
\{\int\limits_0^{\omega _0(q^2,s_0^{D^*})}d\omega
\exp\left(\frac{-s(\omega,q^2)+m_{D^*}^2}{\text{M2}}\right)
\\
\times\left[
%%%%%
-\frac{m_cm_B-2\omega (m_B+\omega)}{\bom^2+
m_c^2-q^2}\phi ^B_-\left(\omega\right)
+\left(\frac{m_cm_B-2\omega\bom-
4\omega^2}{\bom^2+m_c^2-q^2}
-\frac{2\omega +m_B}{\bom}\right)
\phi^B _+\left(\omega\right)
%%%%%%%%%%
\right.
\\
\left.
-\frac{2}{\bom^2+m_c^2-q^2}\left(m_B+2\omega+
\frac{\bom(2\omega \bom+4\omega^2-
m_cm_B)}{\bom^2+m_c^2-q^2}\right)
\text{PhiBpm}(\omega)\right]
\\+\Delta A_{3-0}^{BD^*}(q^2,s_0^{D^*},M^2)
\Bigg \} \,,
\label{eq-A30BDst}
\end{multline}
In (\ref{eq:VBDst})-(\ref{eq-A30BDst}), $\Delta V^{BD^*}$, 
$\Delta A_1^{BD^*}$, $\Delta A_2^{BD^*}$, $\Delta A_{3-0}^{BD^*}$ 
denote the contributions of the  $B$-meson three-particle DA's
collected in the Appendix.

\section{$h_i(w)$  form factors }

In what follows, we use, instead of 
the momentum transfer squared $q^2$, the variable $w$:
\be
w = v\cdot v' = \frac{m_B^2 + 
m_{D^{(\ast)}}^2 - q^2}{2\, m_B\,m_{D^{(*)}}}\,,
\label{eq-w}
\ee 
where $v_\mu = (p+q)_\mu/m_B$ and 
$v'_\mu = p_\mu/m_{\Dst}$  are the four-velocities
of $B$ and $\Dst$. The boundaries of  
the semileptonic region $q^2=0$ and $q^2=(m_B-m_{\Dst})^2$ 
correspond to $w_{max}\simeq 1.589$  
($w_{max}^*\simeq 1.503$) and $w=1$, respectively.

We also switch to the form factors 
adapted to heavy-quark symmetry, defining them as:  
\ba
%\begin{multline}
\frac{\langle D(p)|\bar{c}\gamma_\mu b|\bar B (p+q)
\rangle}{\sqrt{m_B m_D}} &=& 
(v+v')_\mu\, h_+ (w) + (v-v')_\mu\, h_- (w) \,,
\nonumber
\\
\frac{\langle D^\ast (p,\epsilon) |\bar{c}\gamma_\mu b | \bar B (p+q)
\rangle}{\sqrt{m_B m_{D^\ast}}}& = &
\epsilon_{\mu\nu\alpha\beta} \epsilon^{\ast\nu} \ v^\alpha v{'^\beta} \,h_V (w) \,,
\nonumber
\\
\frac{\langle D^*(p,\epsilon)| \bar{c}\gamma_\mu \gamma_5 b|\bar B(p+q)
\rangle}{\sqrt{m_B m_{D^\ast}}} &=& 
\mathnormal{i}\epsilon^{*}_\mu (1+w)h_{A_1}(w) -  
\mathnormal{i}(\epsilon^*\cdot v)\,v_\mu h_{A_2}(w)
\nonumber
\\
&-& \mathnormal{i}(\epsilon^\ast\cdot v)\,v'_\mu h_{A_3}(w)\,.
%\end{multline}
\label{eq-h}
\ea
The functions $h_i(w)$ are related to the initial form factors
defined in (\ref{eq-formfBD}) and (\ref{eq-formfBDst}):
\ba
&&h_{\pm}(w)=\frac1{2\sqrt{r}}\Big[(1\pm r)f^+_{BD}(q^2)+(1\mp r)f^-_{BD}
(q^2)\Big]\,,
 \nonumber\\
&&h_V(w)=\frac{2\sqrt{r^*}}{1+r^*}V^{BD^*}(q^2)\,,~~~~
h_{A_1}(w)=\frac{1+r^*}{\sqrt{r^*}(1+w)}A_1^{BD^*}(q^2)\,,
 \nonumber\\
&&r^*h_{A_2}(w)+h_{A_3}(w)=\frac{2\sqrt{r^*}}{1+r^*}A_2^{BD^*}(q^2)\,,
 \nonumber\\
&&r^*h_{A_2}(w)-h_{A_3}(w)=\frac{4r^*\sqrt{r^*}\big[A_3^{BD^*}(q^2)-
A_0^{BD^*}(q^2)\big]}{1+r^{*2}-2r^*w}
\,,
\label{eq-hirel}
\ea
where $r^{(*)}=m_{\Dst}/m_B$.
We emphasize that the $h_i$ form factors represent  
linear combinations of 
of the initial form factors and no heavy quark limit is involved
in their definitions.  The form factors (\ref{eq-h}) are calculated  
substituting the sum rules (\ref{eq:fplBD})-(\ref{eq-A30BDst})
in the relations (\ref{eq-hirel}).

It is important to check that the form factors predicted from 
the new sum rules obey   
the heavy-quark symmetry relations 
in the limit $m_c,m_b(m_B)\to\infty$.  

For that we need to rescale the  
masses and decay constants  of heavy mesons:
\ba
m_B=m_Q+\bar{\Lambda},~~m_D=\kappa m_Q+\bar{\Lambda},
\label{mBDscal}\\
f_B=\frac{\hat{f}}{\sqrt{m_Q}},~~ 
f_D=\frac{\hat{f}}{\sqrt{\kappa}\sqrt{m_Q}}\,,
\label{fBDscal}
\ea
as well as redefine the effective threshold and Borel parameter
\ba
s_0^D=\kappa^2 m_Q^2+2\kappa m_Q\beta_0,~~
M^2=2\kappa m_Q\tau,
\label{scal} 
\ea
where $m_b\to m_Q$, and the ratio $\kappa=m_c/m_b$. 
Substituting these transformations 
into the sum rules (\ref{eq:fplBD})-(\ref{eq-A30BDst}), 
switching to $h_i$-form factors and taking the $m_Q\to \infty$ 
limit, we readily obtain the usual heavy-quark symmetry relations: 
\ba
h_+(w)= h_{V}(w)=h_{A_1}(w)=h_{A_3}(w)=\xi(w)\,,\nonumber \\
h_-(w)=h_{A_2}(w)=0\,, 
\label{eq-hrel}
\ea
where $\xi(w)$ given by the sum rule: 
\be
\xi(w)=\int\limits_0^{\beta_0/w} d\rho 
\exp\left(\frac{\bar{\Lambda}-\rho w}{\tau}\right)
\Big[\frac{1}{2w}\phi_{-}^B(\rho)+
\big(1-\frac{1}{2w}\big)\phi_{+}^B(\rho) \Big]\,,
\label{eq-IW}
\ee
has to be identified with the IW function.
The $B\to \Dst$ form factors $h_{i}(w)$ 
obtained from the sum rules with finite $m_c$ and $m_B$ 
deviate from the relations (\ref{eq-hrel}), mainly due
to $\sim 1/m_c$ corrections
\footnote {As discussed above,  
in the correlation function we employ the $B$-meson DA's defined 
in HQET, hence, certain $\sim 1/m_b$ corrections are already absent 
in the initial sum rules.}.
Importantly, all three-particle contributions 
to the sum rule for $\xi(w)$ vanish,
being suppressed by at least one power of the 
inverse heavy quark mass. Note also that 
$\xi(w)$ is independent of $\kappa$, as expected.
The sum rule (\ref{eq-IW}) directly relating 
the Isgur-Wise function to the $B$-meson DA's,
is valid near the maximal recoil, in  
the region where the light-cone expansion of the initial 
sum rules can be trusted
\footnote{Note that in the three-point sum rule approach 
based on local OPE the IW function at $w=1$  is also accessible.}.
Considering the formal limit  of (\ref{eq-IW}) at $w\to \infty$ 
we obtain that $\xi(w)$ decreases $\sim 1/w^2$. 
Note that (\ref{eq-IW}) is 
only a tree-level relation,
and in future it will be interesting to investigate the role 
of radiative corrections, which are beyond 
our scope here.

\section{Numerical results }

Turning to the numerical analysis of the sum rules,
we specify the input. The meson masses are
$m_B=5.279$ GeV, $m_D=1.869$ GeV
and $m_{D^*}= 2.01$ GeV \cite{PDG}. 
For the $B$-meson DA's presented in the Appendix
we adopt the same parameters 
as in \cite{KMO2}, in particular,  
the decay constant $f_B=180\pm 30 $  MeV
and the inverse moment 
$\lambda_B(\mbox{1 GeV})= 460 \pm 110$ MeV \cite{BIK} 
(neglecting the evolution of this parameter). 
Both values originate from the two-point sum rules
with $O(\alpha_s)$ accuracy. 
The remaining parameter is $\lambda_E^2 =3/2 \lambda_B^2$ 
specifying the three-particle B-meson DA's modelled 
in \cite{KMO2}. 

%%%%%%%%%%%%%fig2
\begin{figure}[t]
\begin{center}
\includegraphics[width=12cm]{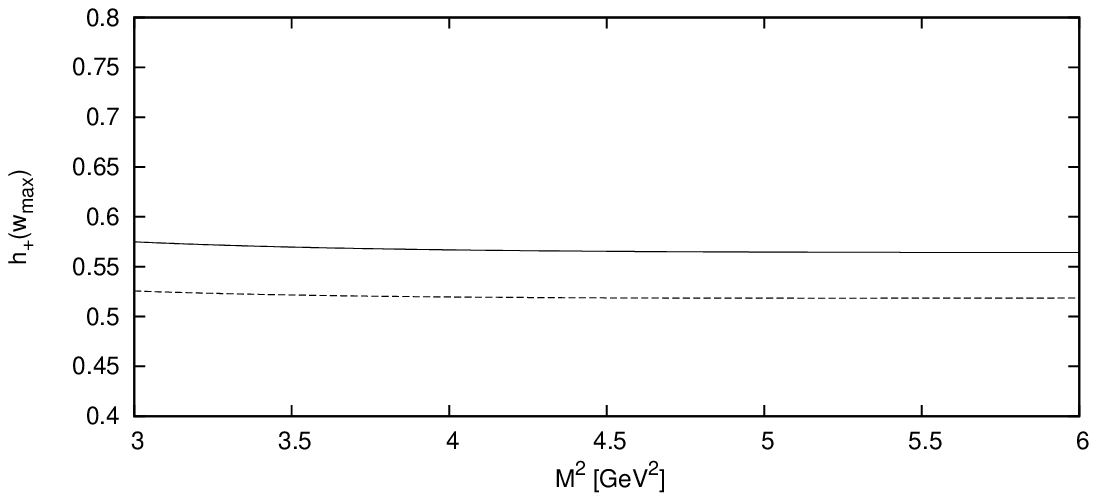}\\
\includegraphics[width=12cm]{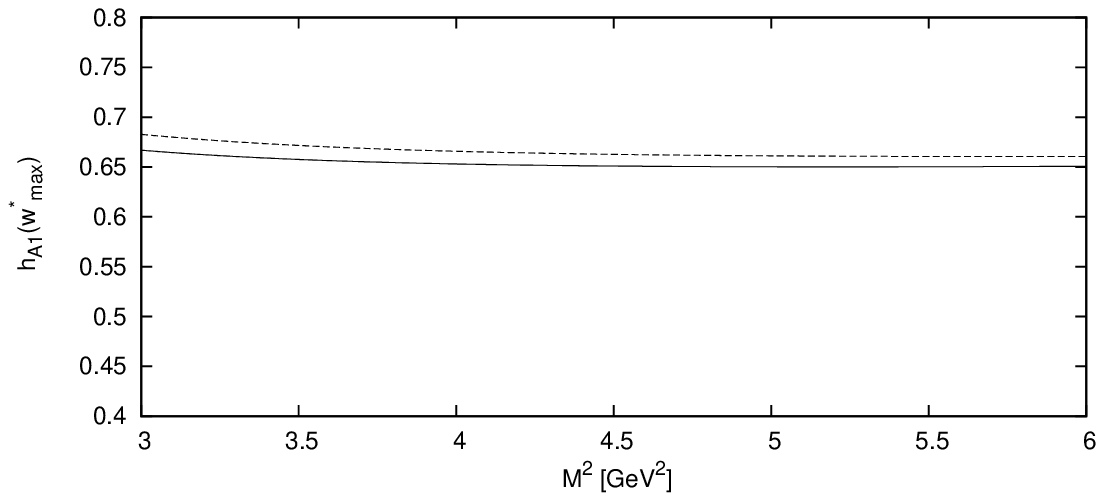}
\end{center}
\caption{ \it Dependence of the $B\to \Dst $ form factors
$h_{+}(w_{max})$ (upper figure) and $h_{A_1}(w_{max}^*)$
(lower figure) on the Borel parameter squared
(solid lines). Dashed lines 
represent the contributions 
of two-particle $B$-meson DA's.}  
\label{fig-Borel}
\end{figure}

As already discussed in section 2,
we use the $\overline{MS}$ mass, with the interval 
$m_c=\overline{m}_c(\overline{m}_c)
=1.25\pm 0.09 $ GeV from \cite{PDG}. 
Note that in our approach there is no need to specify 
the $b$-quark mass value. 
For the decay constants of charmed mesons,
we adopt the intervals determined from the two-point QCD sum rules: 
$f_D= 200\pm 20 $ MeV (see, e.g., \cite{KRWWY,PS,Narison}), 
consistent with the most recent measurement 
\cite{CLEOfD} and  $f_{D^*}=270\pm 30$ MeV (see, e.g., \cite{KRWY2}).

A typical interval used for the Borel mass in 
LCSR for charmed mesons is $M^2=3-6$ GeV$^2$, which we also adopt here.
We then fix the  effective threshold 
$s_0^{\Dst}$ by calculating the 
$\Dst$-meson mass  directly from LCSR,
an approach  frequently used 
in other applications of QCD sum rules.
More specifically, we differentiate 
both parts of each sum rule  
with respect to $1/M^2$ and divide the result 
by the initial sum rule. In this way we obtain 
$s_0^{\Dst}=6.0~(8.0)$ GeV$^2$, with a negligible difference for 
various sum rules. 

To demonstrate an almost perfect stability of LCSR 
with respect to the variation of the Borel parameter,
we plot the form factors $h_{+}(w_{max})$ 
and $h_{A_1}(w_{max}^*)$ of $B\to D$ and 
$B\to D^*$ transitions, respectively, as functions of $M^2$  
in Fig.~\ref{fig-Borel}. All other input parameters
are taken at their central values. From the same figures
it is seen that the contributions of three-particle DA's 
are numerically suppressed.

As argued in sect.~2, the sum rule predictions for $B\to \Dst$ 
form factors can be trusted near the maximal recoil
 $w^{(*)}_{max}$ $(q^2=0)$, where  the light-cone OPE is applicable. 
One more reason to apply the sum rules
at larger $w$ (at smaller $q^2$)  is that 
the upper limits $\omega_0(q^2,s_0^{\Dst})$ in the sum rule 
integrals remain small. Hence, the sum rules
are less dependent on the behavior of the $B$-meson DA's 
at large $\omega$, in particular, on the ``radiative tail''
\cite{NL} not accounted for in our calculation, and are 
to a larger extent sensitive to the inverse moment $\lambda_B$.

Note that the values
of $f_B$ and $f_{\Dst}$ cancel in the ratios of the 
form factors obtained from the new sum rules.
Furthermore, dependence on $\lambda_B$, entering the dominant 
contribution, becomes weaker. Hence, in our approach the 
ratios and slopes of the form factors are in general 
more  accurately predicted, 
than their normalizations.

Let us concentrate our numerical analysis on the 
$B\to D^*$ transition first.
The semileptonic differential rate 
determined by the sum of the three helicity amplitudes squared 
is usually written as:
\be
\frac{d\Gamma(\bar{B}\to D^* l \bar{\nu}_l)}{dw}=
\frac{G_F^2|V_{cb}|^2}{48\pi^3}(m_B-m_D^*)^2 m_{D^*}^3\sqrt{w^2-1}
(1+w)^2 g(w)|{\cal F}(w)|^2\,,
\label{eq:dGammaDst}
\ee
where
\ba
 |{\cal F}(w)|^2=\frac{|h_{A_1}(w)|^2}{g(w)}\Bigg\{
2\Big(\frac{1-2wr^*+r^{*2}}{(1-r^*)^2}\Big)
\nonumber\\
\times\Big[1+\frac{w-1}{w+1}|R_1(w)|^2\Big]+
\Big[1+\frac{w-1}{1-r^*}\big(1-R_2(w)\big)
\Big]^2\Bigg\}
\ea
and $g(w)=1+4w(1-2wr^*+r^{*2})/[(1+w)(1-r^*)^2]$.
This rate is determined by the form factor $h_{A_1}(w)$  
and by the two ratios:
\be
R_1(w)=\frac{h_V(w)}{h_{A_1}(w)}=\Bigg(1-
\frac{q^2}{(m_B+m_{D^*})^2}\Bigg)
\frac{V^{BD^*}(q^2)}{A_1^{BD^*}(q^2)}\,,
\ee
\be
R_2(w)=\frac{r^*h_{A_2}(w)+h_{A_3}(w)}{h_{A_1}(w)}=\Bigg(1-
\frac{q^2}{(m_B+m_{D^*})^2}\Bigg)
\frac{A_2^{BD^*}(q^2)}{A_1^{BD^*}(q^2)} \,.
\ee

The recent BaBar data on $B\to D^* l \nu_l$ differential
rate have been fitted \cite{BaBar1} to the CLN-parameterization 
of the form factors \cite{CLN}, 
based on analyticity and conformal mapping. This parameterization
has the form of a power expansion
in the variable $z=(\sqrt{w+1}-\sqrt{2})/(\sqrt{w+1}+\sqrt{2})$ :
\ba
h_{A_1}(w)=h_{A_1}(1)\big[1-8\rho^2z+(53\rho^2-15)z^2-(231\rho^2-91)z^3\big] \,,
\label{eq-CLNhA}
\\
R_1(w)=R_1(1)-0.12(w-1)+0.05(w-1)^2\,,
\label{eq-CLNR1}
\\
R_2(w)=R_2(1)+0.11(w-1)-0.06(w-1)^2\,,
\label{eq-CLNR2}
\ea
Note that $z\!\ll\! 1 $ in 
the whole semileptonic region  $1<w<w_{max}^*$.
The fit results are \cite{BaBar1}: 
${\cal F} (1) |V_{cb}|=(34.4\pm 0.3\pm 1.1)\times 10^{-3}$,
$\rho^2= 1.191\pm 0.048\pm 0.028 $, $R_1(1)= 1.429\pm 0.061\pm 0.044$, 
$R_2(1)=0.827\pm 0.038\pm 0.022 $. Adopting the 
current average \cite{PDG} from the exclusive 
determinations: 
$|V_{cb}|=(38.6\pm 1.3)\times 10^{-3}$, 
we obtain $h_{A_1}(1)= 0.89\pm 0.04$
\footnote{~This interval agrees within errors with the recent lattice 
QCD result \cite{bernard}:
${\cal F}(1)=h_{A_1}(1)=0.921\pm 0.013\pm 0.020$.}.  The $w$-dependence 
(\ref{eq-CLNhA})-(\ref{eq-CLNR2}) yields:  
$h_{A_1}(w_{max}^*)= 0.52\pm 0.03$, 
$R_1(w_{max}^*)=1.38\pm 0.07 $ 
and $R_2(w_{max}^*)=0.87\pm 0.04$ (adding the statistical 
and systematic errors in quadrature and taking into account
the correlation between the slope and normalization
parameters).

%%%%%%%%%%%%%%%%%%%%%%%%%%%%
\begin{figure}[t]
\begin{center}
\includegraphics[width=12cm]{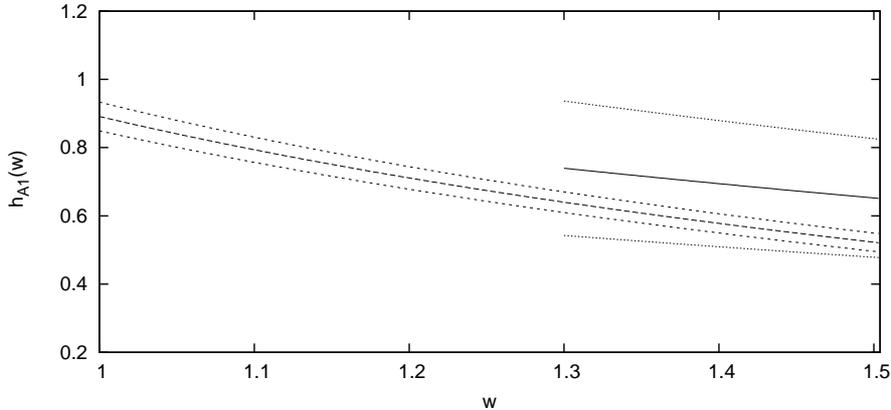}
\end{center}
\caption{ \it Comparison of the
$B\to D^* $ form factor $h_{A_1}(w)$ calculated from 
LCSR at $w>1.3$ (solid),  
with the fit of the BaBar data to 
the CLN parameterization (long-dashed). 
Dotted (short-dashed) 
lines indicate the estimated theoretical uncertainty 
(experimental fit error).}
\label{fig_hA1}
\end{figure}
From the sum rule for $A_1^{BD^*}(q^2=0)$, using the 
third relation in (\ref{eq-hirel}), we obtain
\be
\left[h_{A_1}(w_{max}^*)\right]_{LCSR}= 0.65\pm 0.12
\pm [0.11]_{f_B}\pm [0.07]_{f_{D^*}},
\label{eq:ha1}
\ee
somewhat larger, but still consistent within uncertainties
with the value of this form factor 
extracted from experimental data.
The uncertainty of the sum rule prediction is 
estimated by varying $m_c$,
$\lambda_B$ and Borel parameter $M$ 
within the adopted intervals and adding in quadratures 
the resulting variations of the sum rule result.
The uncertainties caused by $f_B$ and $f_D$ are 
shown separately. Comparing the sum rule predictions
at $w=1.3$ and $w_{max}^*$ with the CLN parameterization
we obtain 
\be  
\rho^2 = 0.81 \pm 0.22\,.
\label{eq:rfit}
\ee
The ratios of $B\to D^*$  form factors 
at maximal recoil
obtained from the combinations of sum rules 
\be
\left [R_1(w_{max}^*)\right]_{LCSR}=1.32\pm 0.04 ,~~~
\left [R_2(w_{max}^*)\right]_{LCSR}=0.91\pm 0.17\,,
\label{eq-R12}
\ee
are in a better agreement with the BaBar data. 

\begin{figure}[t]
\begin{center}
\includegraphics[width=12cm,height=8cm]{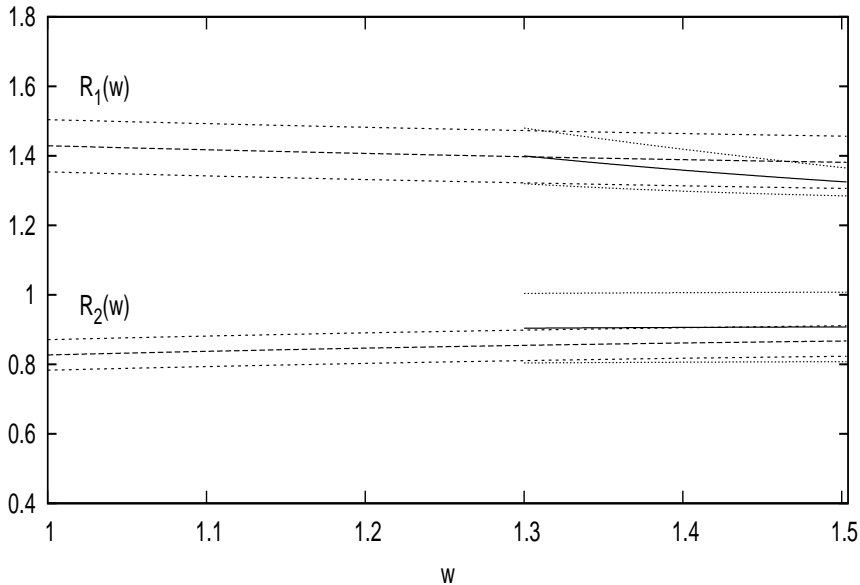}
\end{center}
\caption{ \it 
The ratios of $B\to D^* $ form factors $R_{1}(w)$ (upper)
and $R_{2}(w)$ (lower). The LCSR results (solid 
lines at $w>1.3$) are compared with the fit of the BaBar 
data to the CLN parameterization (long-dashed). 
Dotted (short-dashed) 
lines indicate the estimated theoretical uncertainty 
(experimental fit error).}
\label{fig_R1R2}
\end{figure}

To illustrate our numerical results,
in Fig.~\ref{fig_hA1} and  Fig.~\ref{fig_R1R2} 
we compare the form factor 
$h_{A_1}(w)$ and the ratios $R_{1,2}(w)$, respectively, 
with the BaBar data fitted to CLN parameterizations.  
%%%%%%%%%%%%%%%%fig4
%%%%%%%%%%%%%%%%%%%%%%%%%%%

Furthermore, we present the numerical predictions 
for $B\to D$ form factors 
comparing them with the latest measurement 
by BaBar collaboration \cite{BaBar2}.
In the differential rate of $\bar{B}\to D l \bar{\nu}_l$ 
\be
\frac{d\Gamma(\bar{B}\to D l \bar{\nu}_l)}{dw}=
\frac{G_F^2|V_{cb}|^2}{48\pi^3}(m_B+m_D)^2m_D^3(w^2-1)^{3/2}
|{\cal G}(w)|^2\,.
\label{eq:dGammaD}
\ee
the two form factors $h_{\pm}$ are combined within a single function: 
\be
{\cal G}(w)=  h_+ (w)-\frac{1-r}{1+r}\,h_-(w)\,. 
\label{eq:Gw}
\ee
In \cite{BaBar2} the CLN-parameterization \cite{CLN} 
for this form factor was used: 
\be
{\cal G}(w)=
{\cal G}(1)\{1-8\rho_D^2z +(51\rho_D^2-10)z^2-
(252\rho_D^2-84)z^3\}
\label{eq:calG}
\ee 
yielding the following fitted values:
$|V_{cb}|{\cal G}(1)=
(43.0\pm 1.9\pm 1.4)\times 10^{-3}$, 
$\rho_D^2=1.20\pm 0.09\pm 0.04$.
With  the same value $|V_{cb}|$ as used above, 
we obtain $ {\cal G}(1)= 1.11 \pm 0.07$ and 
$ {\cal G}(w_{max})= 0.60 \pm 0.02$.  

The sum rules for $f^+(0)$ and $[f^+(0)+
f^-(0)]$, combined with the first relation 
in (\ref{eq-hirel}) and with (\ref{eq:Gw}) yield:  
\be
\left [{\cal G}(w_{max})\right]_{LCSR}= 0.61\pm 0.11 
\pm [0.10]_{f_B}\pm [0.07]_{f_{D}},
\ee
\be  
\rho^2_D = 1.15 \pm 0.15\,,
\label{eq:Vslopefit}
\ee
in a reasonable agreement with the experimental results.
The sum rule prediction for ${\cal G}(w)$ in the region 
$1.3<w<w_{max}$ is plotted in Fig.~5, compared with 
the new BaBar data.

%%%%%%%%%%%%%%%%fig5
\begin{figure}[t]
\begin{center}
\includegraphics[width=12cm]{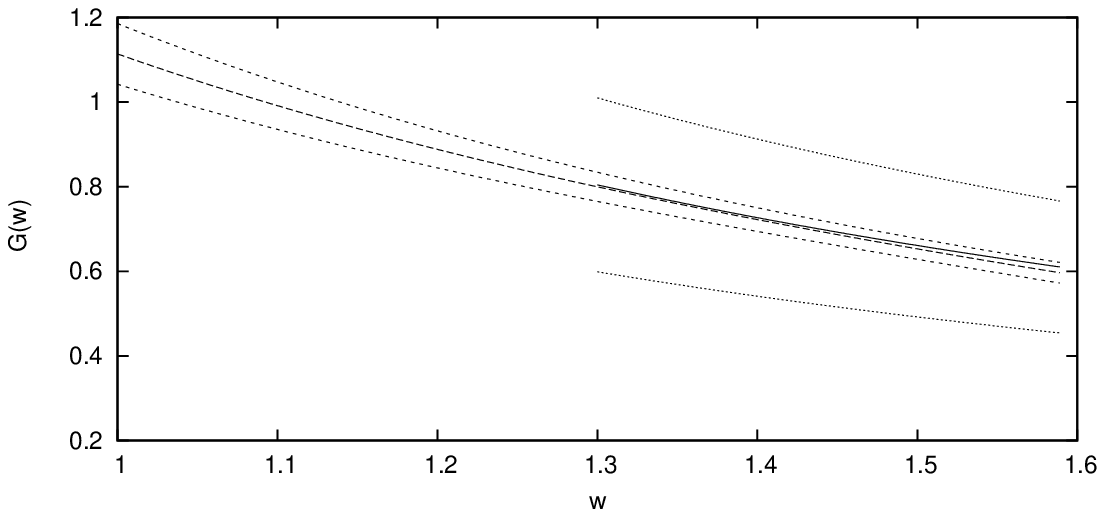}
\end{center}
\caption{ \it 
The combination of $B\to D $ form factors ${\cal G}(w)$ 
calculated from LCSR at $w>1.3$ (solid), compared 
with the fits of the BaBar data to 
the CLN parameterization (long-dashed). 
The  dotted (short-dashed) 
lines indicate the theoretical uncertainty 
(experimental fit error).}
\label{fig_calV}
\end{figure}
%%%%%%%%%%%%%%%%%%%%%%%%%%%

Finally, it is instructive to compare the numerical result 
for $\xi(w)$ inferred from the limiting sum rule 
(\ref{eq-IW}) using the same input  parameters
as for the finite mass sum rules and rescaling them
according to (\ref{mBDscal})  and (\ref{scal}). 
We obtain for the central values of the input 
$\xi(w_{max})=0.72$, in the ballpark of three-point sum rule
predictions (see e.g.,~\cite{Neubert}). On the other hand, 
comparison with the corresponding central value 
of $h_+(w_{max})=0.56$ reveals 
a substantial deviation from the heavy-quark symmetry relations  
(\ref{eq-hrel}) in the region of maximal recoil, and a
somewhat smaller deviation for  $h_{A_1}(w_{max}^*)$.
In order to illustrate the transition of  
this form factor from its central value (\ref{eq:ha1}) 
at finite $m_c$  to the
heavy-quark limit $\xi(w_{max}^*)=0.73$, in Fig.~6. we plot the 
dependence of the LCSR for $h_{A_1}(w_{max}^*)$ on
$m_c=\kappa m_Q$ at $m_Q\to \infty$.
The symmetry violation for the remaining $B\to D^*$ form factors 
is determined by $R_{1,2}(w_{max})\neq 1$.  
 %%%%%%%%%%%%%%%%fig6
\begin{figure}[h]
\begin{center}
\includegraphics[width=10cm]{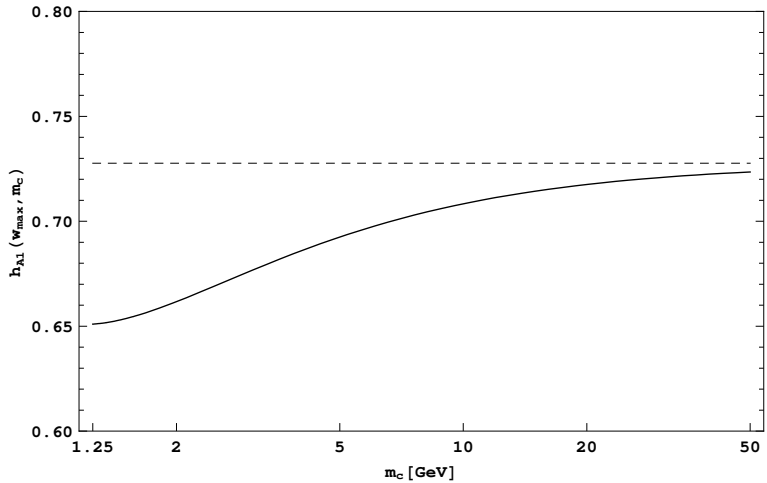}
\end{center}
\caption{ \it Dependence of the form factor $h_{A_1}(w_{max}^*)$ on $m_c$  
(solid), compared with the heavy-quark limit (dashed), 
at central values of the input parameters.}
\end{figure}

\section{Conclusion}

In this paper we present the first exploratory 
application of QCD light-cone sum rules 
to the $B\to \Dst$ form factors, 
a traditional testing ground of HQET. 
We use the recently developed version of LCSR involving 
$B$-meson DA's. These sum rules are valid 
at large recoils $w\sim w_{max}$, complementing 
the rich theoretical knowledge of $B\to \Dst$ 
form  factors at the zero-recoil point.
The sum rules are obtained at the finite $c$-quark mass, 
allowing one to investigate the deviations from HQET.
The quark-hadron duality in the charmed meson channel
employed in our approach is better understood and 
presumably introduces
a smaller systematic uncertainty than the duality ansatz in 
double dispersion relations used for three-point sum rules.
Moreover, it is possible,
by  combining the sum rules obtained here and in \cite{KMO2},   
to calculate  the ratios of $B\to \pi,\rho$ and $B\to \Dst$ 
form factors, employing the same approach and input
and to extract the ratio $|V_{ub}|/|V_{cb}|$.

Within limited accuracy 
of our calculation, we observe a reasonable agreement 
with the experimental data,
encouraging further development of the LCSR 
approach with $B$-meson DA's for $b\to c$ exclusive transitions.
It is possible e.g., to calculate  the form factors of $B$-meson 
transitions  to the excited $D$-meson states.  

In order to turn the sum rules suggested here 
into a truly competitive tool for the 
theoretical analysis of $B\to \Dst$ form factors, 
a better knowledge of B-meson DA's and heavy meson decay constants 
is desirable. Importantly,   
one also has to calculate the gluon radiative corrections to the 
correlation function including the renormalization of $B$-meson DA's, 
a task that we postpone to a future study.\\[2mm]

\noindent {\bf Acknowledgements}\\[1mm]
We are grateful to 
Thorsten Feldmann and Nils Offen for useful discussions.
This work was supported by the Deutsche Forschungsgemeinschaft 
under the  contract No. KH205/1-2.

\section*{Appendix}

\subsection*{B-meson DA's}

We use the following definitions of the two-particle 
$B$-meson DA's:
\begin{eqnarray}
&&
\langle 0|\bar{q}_{2\alpha}(x)[x,0] h_{v\beta}(0)
|\bar{B}_v\rangle
= -\frac{if_B m_B}{4}\int\limits _0^\infty 
d\omega e^{-i\omega v\cdot x} 
\nonumber \\ 
&&\times
\left [(1 +\DS v)
\left \{ \phi^B_+(\omega) -
\frac{\phi_+^B(\omega) -\phi_-^B(\omega)}{2 v\cdot x}\DS x \right \}\gamma_5\right]_{\beta\alpha}\,.
\label{eq-BDAdef}
\end{eqnarray}
The DA's 
$\phi_{+}^B(\omega)$ and $\phi_{-}^B(\omega)$  
are normalized with $\int_0^\infty d\omega\phi^B_{\pm}(\omega)=1$,
the variable $\omega>0$ being the plus component of the 
spectator-quark momentum in the $B$ meson\footnote{Note that the 
integrals over $\phi^B_{\pm}$ enter the sum rules with upper
bounds, hence the ``radiative tail'' emerging \cite{NL} 
after taking into account nontrivial renormalization 
properties of these functions is not important.}.   

For the three-particle DA's the definition
\cite{Kawamura} is employed:
\begin{eqnarray}
&&\langle 0|\bar{q_2}_\alpha(x) G_{\lambda\rho}(ux) 
h_{v\beta}(0)|\bar{B}^0(v)\rangle=
\frac{f_Bm_B}{4}\int\limits_0^\infty d\omega
\int\limits_0^\infty d\xi\,  e^{-i(\omega+u\xi) v\cdot x} 
\nonumber \\ 
&&\times \Bigg [(1 +\DS v) \Bigg \{ (v_\lambda\gamma_\rho-v_\rho\gamma_\lambda)
\Big(\Psi_A(\omega,\xi)-\Psi_V(\omega,\xi)\Big)
-i\sigma_{\lambda\rho}\Psi_V(\omega,\xi)
\nonumber\\
&&-\left(\frac{x_\lambda v_\rho-x_\rho v_\lambda}{v\cdot x}\right)X_A(\omega,\xi)
+\left(\frac{x_\lambda \gamma_\rho-x_\rho \gamma_\lambda}{v\cdot x}\right)Y_A(\omega,\xi)\Bigg\}\gamma_5\Bigg]_{\beta\alpha}\,.
\label{eq-B3DAdef}
\end{eqnarray}
In the above, the path-ordered gauge factors are omitted for
brevity. The DA's $\Psi_{V}$,$\Psi_{A}$, $X_A$ and $Y_A$ depend 
 on the two variables $\omega>0$ and $\xi>0$ being, respectively, 
the plus components of the light-quark and gluon momenta 
in the $B$ meson. 

For numerical analysis we use the simple exponential 
model suggested in \cite{GN} for two-particle DA's
\begin{eqnarray}
\phi_+^B(\omega) & = & \dfrac{\omega}{\omega_0^2}\,e^{-\frac{\omega}{\omega_0}}\,,\nonumber\\
\phi_-^B(\omega) & = & \dfrac{1}{\omega_0}\,e^{-\frac{\omega}{\omega_0}}\,,
\label{eq-GN}
\end{eqnarray}
where the inverse moment $\lambda_B$ defined as 
$1/\lambda_B=\int_0^\infty \frac{d\omega}{\omega} \phi_+^B(\omega)$ 
is equal to $\omega_0$.

For the three-particle DA's we use
the exponential ansatz suggested in \cite{KMO2}:
\begin{eqnarray}
\Psi_A(\omega,\,\xi)& =& \Psi_V(\omega,\,\xi) \,=\, 
\dfrac{\lambda_E^2 }{6\omega_0^4}\,\xi^2 e^{-(\omega\,+\,\xi)/\omega_0}\,,
\nonumber\\
X_A(\omega,\,\xi)& = & \dfrac{\lambda_E^2 }{6\omega_0^4}\,
\xi(2\omega-\xi)\,e^{-(\omega\,+\,\xi)/\omega_0}\,,\nonumber\\
Y_A(\omega,\,\xi)& =&  -\dfrac{\lambda_E^2 }{24\omega_0^4}\,
\xi(7\omega_0-13\omega+3\xi)e^{-(\omega\,+\,\xi)/\omega_0}\,.
\label{eq-3partexp}
\end{eqnarray}

\subsection*{Contributions of three-particle 
DA's to LCSR}
Here we present the contributions
of three-particle DA's to LCSR, expressed in a generic form: 
\begin{multline}
\Delta F(q^2,s_0^{\Dst},M^2)\: =
\:\int\limits_0^{\omega_0(q^2,s_0^{\Dst})/m_B}\, d\sigma 
\exp\left(\frac{-s(\sigma m_B,q^2)+m_{\Dst}^2}{\text{M2}}\right) \\
\times\left( -I^{(F)}_1(\sigma) +\frac{I^{(F)}_2(\sigma)}{M^2}-
\frac{I^{(F)}_3(\sigma)}{2M^4}\right)
\\
+ \frac{e^{(-s_0^{\Dst}+m_{\Dst}^2)/M^2}}{m_B^2}\Bigg\{\eta(\sigma)\Bigg[ I_2^{(F)}(\sigma)\\
-\frac12\left( \frac{1}{M^2} 
+\frac{1}{m_B^2}\frac{d\eta(\sigma)}{d\sigma}\right)I_3^{(F)}(\sigma)
-\frac{\eta(\sigma)}{2m_B^2}\frac{dI_3^{(F)}(\sigma)}{d\sigma}\Bigg]\Bigg\}\Bigg 
|_{\sigma=\omega_0/m_B}\,,
\end{multline}
where
\be
\Delta F=\Big \{\Delta f_{BD}^+,
-\Delta f_{BD}^{\pm},
 \frac{\Delta V^{BD^*}}{m_B}, 
\frac{\Delta A_1^{BD^*}}{m_B}, 
\frac{\Delta A_2^{BD^*}}{m_B}, 
\frac{\Delta A_{3-0}^{BD^*}}{m_B}\Big \} 
\ee
and the following notation is used: 
\be
\eta(\sigma)= \left(1+\frac{m_c^2-q^2 }{\bar{\sigma}^2 m_B^2}\right)^{-1}\,.
\ee 
The integrals
over the three-particle DA's multiplying the inverse powers 
of the Borel parameter $1/M^{2(n-1)}$ with $n=1,2,3$
are defined as:
\begin{multline}
I^{(F)}_n(\sigma)= \frac{1}{\bar{\sigma}^n}\int\limits_0^{\sigma m_B} d\omega 
\int\limits_{\sigma m_B-\omega}^{\infty} 
\frac{d\xi}\xi \Bigg[ C^{(F,\Psi A)}_n(\sigma,u,q^2)\Psi_A(\omega,\xi)\\
+C^{(F,\Psi V)}_n(\sigma,u,q^2)\Psi_V(\omega,\xi) 
\\
+C^{(F,XA)}_n(\sigma,u,q^2)\overline{X}_A(\omega,\xi)
+C^{(F,YA)}_n(\sigma,u,q^2)\overline{Y}_A(\omega,\xi)
\Bigg]\Bigg|_{u=(\sigma m_B -\omega)/\xi }
\label{intn}
\end{multline}
where:
$$\overline{X}_A(\omega,\xi)=\int\limits_0^\omega d\tau X_A(\tau,\xi),
~~\overline{Y}_A(\omega,\xi)=\int\limits_0^\omega d\tau Y_A(\tau,\xi).$$

The nonvanishing coefficients entering Eq.~(\ref{intn}) are: 
%%%%%%%%%%%%%%%%f+BD
%\begin{multline}
\ba
C_1^{(f^+_{BD},\Psi A)}= -2\frac{1-u}{\text{mB} \text{ubar}}\,,
\nonumber 
\\
C_2^{(f^+_{BD},\Psi A)}= 
\text{mB}\text{ubar}(4 u-1) +3 \text{mc}-
2 \frac{\text{mc}^2-q^2}{\text{mB} \text{ubar}}(1-u)\,,
\nonumber 
\\
C_1^{(f^+_{BD},\Psi V)}= \frac{2 (1-u)}{\text{mB} \text{ubar}}\,,
\nonumber 
\\
C_2^{(f^+_{BD},\Psi V)}=
 \text{mB}\text{ubar}(2 u+1)+3\text{mc}
    +2 \frac{\text{mc}^2-q^2}{\text{mB}\text{ubar}} (1-u)
\,,
\nonumber 
\\
C_2^{(f^+_{BD},XA)}= 1 - 2 u-\frac{2 \text{mc}}{\text{mB} \text{ubar}}\,,
\nonumber 
\\
C_3^{(f^+_{BD},XA)}=
2\Big(
\text{mc}\text{mB}\text{ubar}
+\text{mB}^2 \text{ubar}^2 (1-2 u)
-\frac{\text{mc}(\text{mc}^2-q^2)}{\text{mB}\text{ubar}} 
\nonumber 
\\ 
-\left(\text{mc}^2+q^2\right)(1-2 u)\Big)\,,
\nonumber 
\\
C_3^{(f^+_{BD},YA)}= -12 \text{mc} \big(\text{mB} \text{ubar}
-\text{mc} (1-2 u)\big)\,,
%\end{multline}
\ea

%%%%%%%%%%%%%%%%%%%%fBD pm
%\begin{multline}
\ba
C_1^{(f_{BD}^{\pm}, \Psi A)}= -2\frac{1-u}{\text{mB} \text{ubar}}\,,
\nonumber \\
C_2^{(f_{BD}^{\pm}, \Psi A)}=
   - \Big(\text{mB}[1+(1-4 u)\text{ubar}+2 u]-3 \text{mc} 
%\nonumber \\
+2 \frac{\text{mc}^2- q^2}{\text{mB} \text{ubar}}(1-u)\Big)\,,
\nonumber \\
C_1^{(f_{BD}^{\pm}, \Psi V)}= 2\frac{1-u}{\text{mB} \text{ubar}}\,,
\nonumber \\
C_2^{( f_{BD}^{\pm}, \Psi V)}= 
\text{mB}[1+(1+2u)\text{ubar}-4u]+3 \text{mc}
%\nonumber \\
+2 \frac{\text{mc}^2-q^2}{\text{mB} \text{ubar}}(1-u)\,,
\nonumber \\
C_2^{(f_{BD}^{\pm},XA)}=-\frac{
\text{mB} (1+\sigma) (1-2 u)+2 \text{mc}}{\text{mB}\text{ubar}}\,,
\nonumber \\
C_3^{(f_{BD}^{\pm},XA)}=
-2\Big(\text{mB}^2\sigma  \text{ubar} (1-2 u) 
+\text{mc}\text{mB}(1+\sigma) \nonumber \\
+\frac{\left[\text{mc}^2(1+\text{ubar})
-q^2\sigma\right]}{\text{ubar}}(1-2 u)
%\nonumber \\
+\frac{m_c(\text{mc}^2-q^2)}{\text{mB} \text{ubar}} 
\Big)\,,
\nonumber \\[1.5mm]
C_3^{(f_{BD}^{\pm},YA)}= 
12 \text{mc} \Big(\text{mB}\sigma+\text{mc} (1-2 u)\Big)\,,
%\end{multline}
\ea

%%%%%%%%%%%%%%%%%%%%V 
\ba
C_2^{( V^{BD^*}\!\!,\Psi A)}&=&-\frac{1-2 u}{m_B}\,,~~
%%
%\nonumber \\
%%
C_2^{( V^{BD^*}\!\!,\Psi V)}=-\frac{1}{m_B}\,,~~
%%
%\nonumber \\
%%
C_2^{(V^{BD^*}\!\!,XA)}=2\frac{1-2u}{m_B^2\text{ubar}}\,,~~
\nonumber \\
C_3^{(V^{BD^*}\!\!,XA)}&=& 
2 \Big(\text{ubar}(1-2 u) +2 \frac{m_c}{m_B} +
\frac{m_c^2 -q^2}{m_B^2\text{ubar}} (1-2 u)\Big)\,,
\nonumber \\
C_3^{(V^{BD^*}\!\!,YA)}&=&-4\frac{ m_c}{ m_B}\,,
\ea

%%%%%%%%%%%%%%%%%%%%A1
\ba
C_1^{( A_1^{BD^*}\!\!,\Psi A)}&=&-\frac{1-2 u}{m_B^2\text{ubar}}\,,
\nonumber \\
C_2^{( A_1^{BD^*}\!\!,\Psi A)}&=&-\text{ubar}(1-2 u)
+2 \frac{m_c}{m_B} 
-\frac{m_c^2-q^2}{m_B^2\text{ubar}}(1-2 u) \,,
\nonumber \\
C_1^{( A_1^{BD^*}\!\!, \Psi V)}&=&-\frac{1}{m_B^2\text{ubar}}\,,
%%
%\nonumber \\
%%
~~~C_2^{( A_1^{BD^*}\!\!, \Psi V)}= - \Big(\text{ubar}
+2 \frac{m_c}{m_B} +\frac{m_c^2-q^2}{m_B^2\text{ubar}}\Big)
\,,
\nonumber \\
C_1^{(A_1^{BD^*}\!\!,XA)}&=&2\frac{1-2u}{m_B^3\text{ubar}^2}\,,
\nonumber \\
C_2^{(A_1^{BD^*}\!\!,XA)}&=&\frac{2}{m_B}\Big( 
1+2 \frac{m_c^2-q^2}{m_B^2\text{ubar}^2}\Big)(1-2u)
\,,
\nonumber \\
C_3^{(A_1^{BD^*}\!\!,XA)}&=& 2\Big(m_B\text{ubar}^2-
2\frac{m_c^2+q^2}{m_B}+ \frac{(m_c^2-q^2)^2}{m_B^3\text{ubar}^2}\Big)(1-2u)\,,
\nonumber \\
C_2^{(A_1^{BD^*}\!\!,YA)}&=& -\frac{4}{m_B} \Big(
1-2 u+\frac{m_c}{m_B\text{ubar}}\Big)\,,
\nonumber \\
C_3^{(A_1^{BD^*}\!\!,YA)}&=&-4 m_c \Big(\text{ubar} 
-2 \frac{m_c}{m_B}(1-2 u)+
\frac{m_c^2-q^2}{m_B^2\text{ubar}}\Big)
\,,
\ea

%%%%%%%%%%%%%%%%%%%%A2
\ba
C_2^{(A_2^{BD^*},\Psi A)}= -\left(1+2u+2\sigma(1-2u)-\frac{4 m_c}{m_B}\right)\,,
\nonumber \\
C_2^{( A_2^{BD^*}, \Psi V)}= -\left(1+2\sigma-4 u +\frac{4 m_c}{m_B}\right)\,,
\nonumber \\
C_2^{(A_2^{BD^*},XA)}=-\frac{2\sigma}{m_B\text{ubar}} (1-2 u)
\,,
\nonumber \\
C_3^{(A_2^{BD^*},XA)}= 2\Big(m_B \text{ubar}(2\text{ubar}-1)(1-2u)-2 m_c 
 \nonumber \\
-\frac{[m_c^2(2 \text{ubar}+1)+q^2(2\text{ubar}-1) ]}{m_B\text{ubar}} 
(1-2 u) \Big)\,,
\nonumber\\
C_3^{(A_2^{BD^*},YA)}=4 \Big(
2 m_B \sigma \text{ubar}(1-2 u)-m_c (1-4 \sigma)\Big)\,,
\ea

%%%%%%%%%%%%%%%%%%%%%%%%%%%%%%%%%%%A3-A0
\ba
 C_2^{(A_3^{BD^*}-A_0^{BD^*}, \Psi A)} = 
2(1-2 u) \bar{\sigma }-1+6 u+\frac{4 m_c}{m_B}
\,,
\nonumber \\
 C_2^{(A_3^{BD^*}-A_0^{BD^*}, \Psi V)} = 
2 \bar{\sigma }-1-4 u-\frac{4 m_c}{m_B}
\,,
\nonumber \\
 C_2^{(A_3^{BD^*}-A_0^{BD^*}, XA)} = -\frac{2 (2 u-1) \left(\bar{\sigma }-3\right)}{\bar{\sigma } m_B}\,,
\nonumber \\
C_3^{(A_3^{BD^*}-A_0^{BD^*}, XA)} = 
2\Big( m_B \bar{\sigma }(2\bar{\sigma }-3)(1-2u)
+2 m_c
\nonumber \\
-\frac{m_c^2(2\bar{\sigma}+3) +q^2 (2\bar{\sigma}-3)}{\bar{\sigma } m_B}
(1-2 u)\Big)\,,
\nonumber \\
C_3^{(A_3^{BD^*}-A_0^{BD^*}, YA)} = 4 \Big(2m_B 
\left[\bar{\sigma }(\sigma+2)-2\right](1-2 u)  
+m_c\left(1+4 \sigma \right) \Big)\,.
\ea

\end{document}